\documentclass{PoS}

\PoS{PoS(HEP2005)110}

\title{Particle production at HERA}

\ShortTitle{Particle production at HERA}

\author{\speaker{Chuanlei Liu}\thanks{On behalf of the H1 and ZEUS Collaborations.}\\
        McGill University, Montreal, Canada\\
        E-mail: \email{liuc@mail.desy.de}}

\abstract{Recent results on the properties of the hadronic final state at the HERA collider are presented. Strangeness production and correlations have been studied in the laboratory system for deep inelastic scattering (DIS). Inclusive $K^0_S$, $\Lambda$ and $\bar{\Lambda}$ production has been measured and compared to Monte Carlo model calculations. No significant $\Lambda$ to $\bar{\Lambda}$ production asymmetry was observed and the measured $\Lambda (\bar{\Lambda})$ polarisations are consistent with zero. Bose-Einstein correlations between charged and neutral kaons have been studied and the obtained results are compared with the LEP measurements. A search for heavy particles was performed in the photoproduction process. The cross section for anti-deuteron production has been measured at a mean center-of-mass energy of $W_{\gamma p}$ = 200 GeV. The azimuthal angle distribution of hadrons in DIS has also been studied in the hadronic center-of-mass system. The measurements of the azimuthal asymmetry are consistent with the perturbative QCD predictions.}

\FullConference{International Europhysics Conference on High Energy Physics\\
		 July 21st - 27th 2005\\
		 Lisboa, Portugal}

\begin{document}

\section{Introduction}
Recent results on a selection of topics concerning the hadronic final state are reviewed. The data samples used for these analyses were taken by H1 or ZEUS Collaborations at the HERA $ep$ collider during the running period 1996-2000. The corresponding center-of-mass energy of the $ep$ collision is 300 GeV in 1996-1997 and 318 GeV in 1998-2000. The large data samples provide rich and valuable information on particle production. 
\section{Strangeness production and correlations}
Measurements of $K^0_S$ and $\Lambda (\bar{\Lambda})$\footnote{Hereafter, both $\Lambda$ and $\bar{\Lambda}$ are referred to as $\Lambda$, unless an explicit comparison is made.}  can be used as a tool to test strange quark production and fragmentation implemented in Monte Carlo (MC) models. Furthermore, the enhancement in the production of identical kaons with similar momenta, known as Bose-Einstein effect, provides a way to study the space-time structure of their sources. \\
Inclusive neutral strangeness $K^0_S$ and $\Lambda$ productions were measured by ZEUS Collaboration using an integrated luminosity of 120 $pb^{-1}$ for $\mathrm{Q}^2 > 25$ $\mathrm{GeV}^2$. The $K^0_S$ meson and $\Lambda$ baryon were reconstructed by their decay modes $K^0_S \rightarrow \pi^+\pi^-$ and $\Lambda \rightarrow p\pi^-$. The $\Lambda$ polarisation was measured via the angular distribution of its decay product in the $\Lambda$ rest frame. Data were corrected for detector effects by using the ARIADNE MC model with the strangeness suppression factor $\lambda_s = 0.3$.
\begin{figure}[h]
\begin{center}
\includegraphics[width=3.0in, height=3.0in]{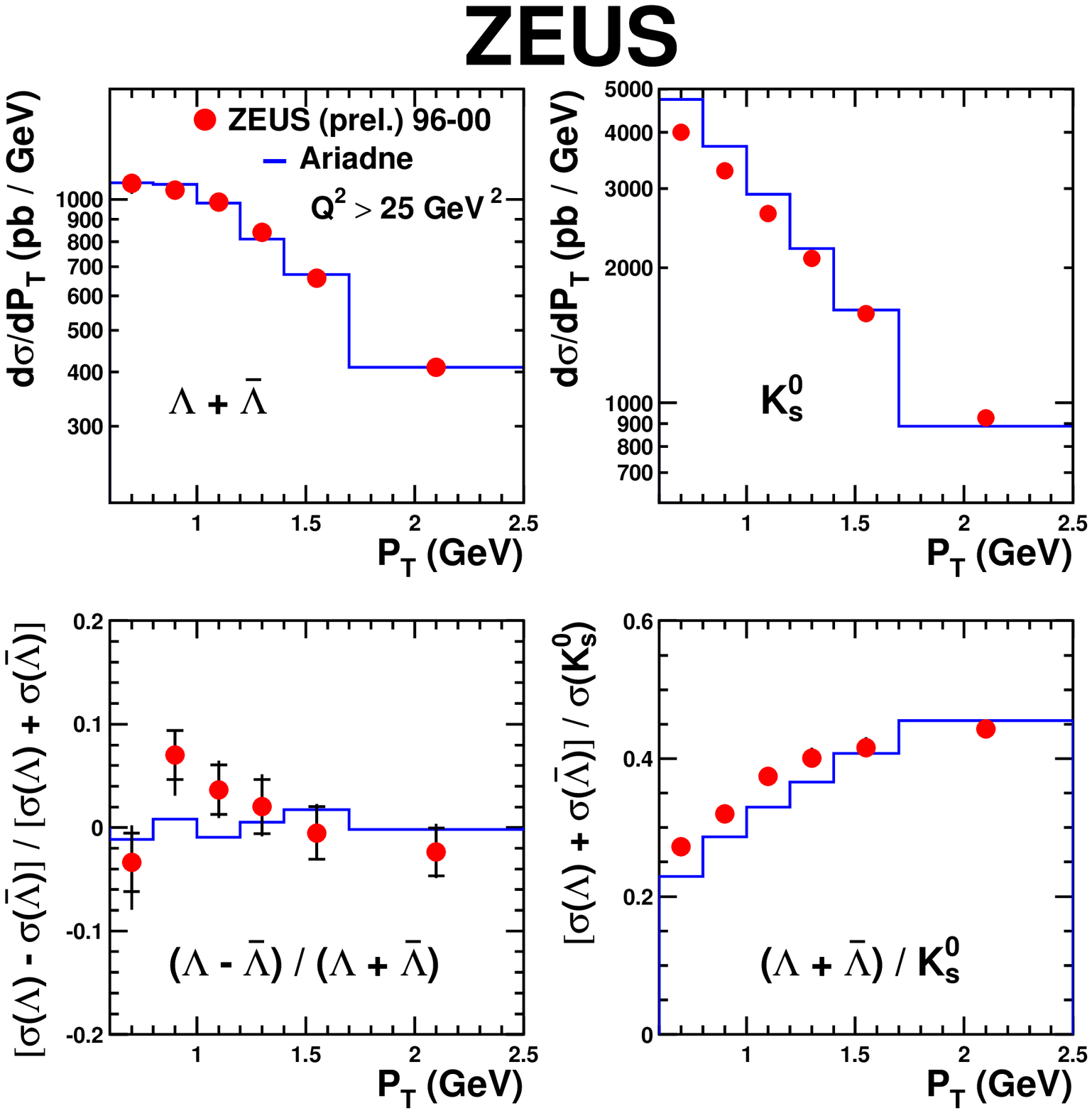}
\includegraphics[width=2.7in, height=2.8in]{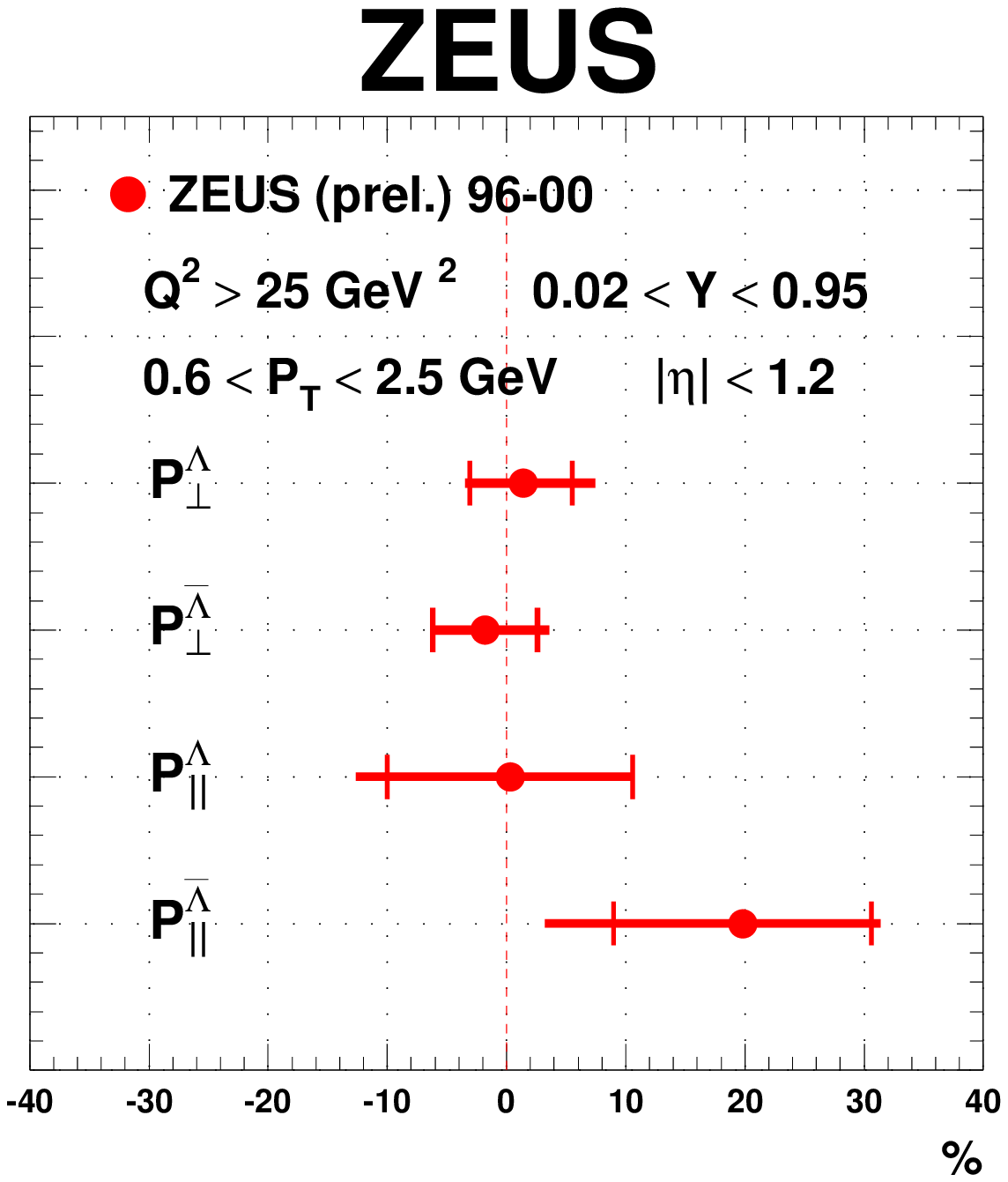}
\end{center}
\caption{$K^0_S$ and $\Lambda$ cross sections and their ratios as a function of $p_T$ in the kinematic region given by $\mathrm{Q}^2 > 25\, \mathrm{GeV}^2$ and $0.02 < y < 0.95 $ (left plot). The measured $\Lambda$ and $\bar{\Lambda}$ polarisations (right plot).}
\label{fig1}
\end{figure} \\
The differential $\Lambda$ and $K^0_S$ cross sections, $\Lambda/\bar{\Lambda}$ asymmetry and $\Lambda/K^0_S$ ratio as a function of $p_T$ are shown in Fig. 1 (left). ARIADNE generally describes the $\Lambda$ cross section well but overestimates the $K^0_S$ cross section at lower $p_T$. Because of this, the $\Lambda/K^0_S$ ratio is also underestimated by the MC in the same $p_T$ region. Detailed MC studies are necessary in order to understand how the MC parameters such as $\lambda_s$ influence the production. No significant $\Lambda/\bar{\Lambda}$ asymmetry was observed. The results on the $\Lambda$ polarisation are presented in Fig. 1 (right). The transverse and longitudinal polarisations of $\Lambda$ were measured to be consistent with zero. 

Bose-Einstein correlations (BEC) between charged and neutral kaons have been measured for the first time in the $\mathrm{Q}^2$ region 2 $<\mathrm{Q}^2 <$ 15000 $\mathrm{GeV}^2$ with the ZEUS detector. A previous ZEUS study of charged pions shows that the BEC are independent of $\mathrm{Q}^2$\cite{zeus}. The $dE/dx$ information was used for the charged kaon identification while the neutral kaon was reconstructed with the decay channel $K^0_S \rightarrow \pi^+\pi^-$. The correlation function is measured using the double ratio method
\begin{center}
$R(Q_{12})=\frac{P(Q_{12})^{data}}{P_{mix}(Q_{12})^{data}}/\frac{P(Q_{12})^{MC, no BEC}}{P_{mix}(Q_{12})^{MC, no BEC}}$,
\end{center}
where the reference sample $P_{mix}$ is the so-called mixed event sample which contains pairs of bosons from different events and $Q_{12}$ is the four momentum difference between two particles. The MC event sample did not include BEC. The particle source information was extracted by fitting the correlation function with the standard Goldhaber-like function\cite{gold}:
\begin{center}
$R(Q_{12}) = \alpha(1+\delta Q_{12}) (1+ \lambda e^{-Q^2_{12} r^2})$.
\end{center}
where parameter $\lambda$ and $r$ are the strength of correlation and the radius of the boson emission source, respectively.
\begin{figure}[h]
\begin{center}
\includegraphics[width=2.4in, height=2.3in]{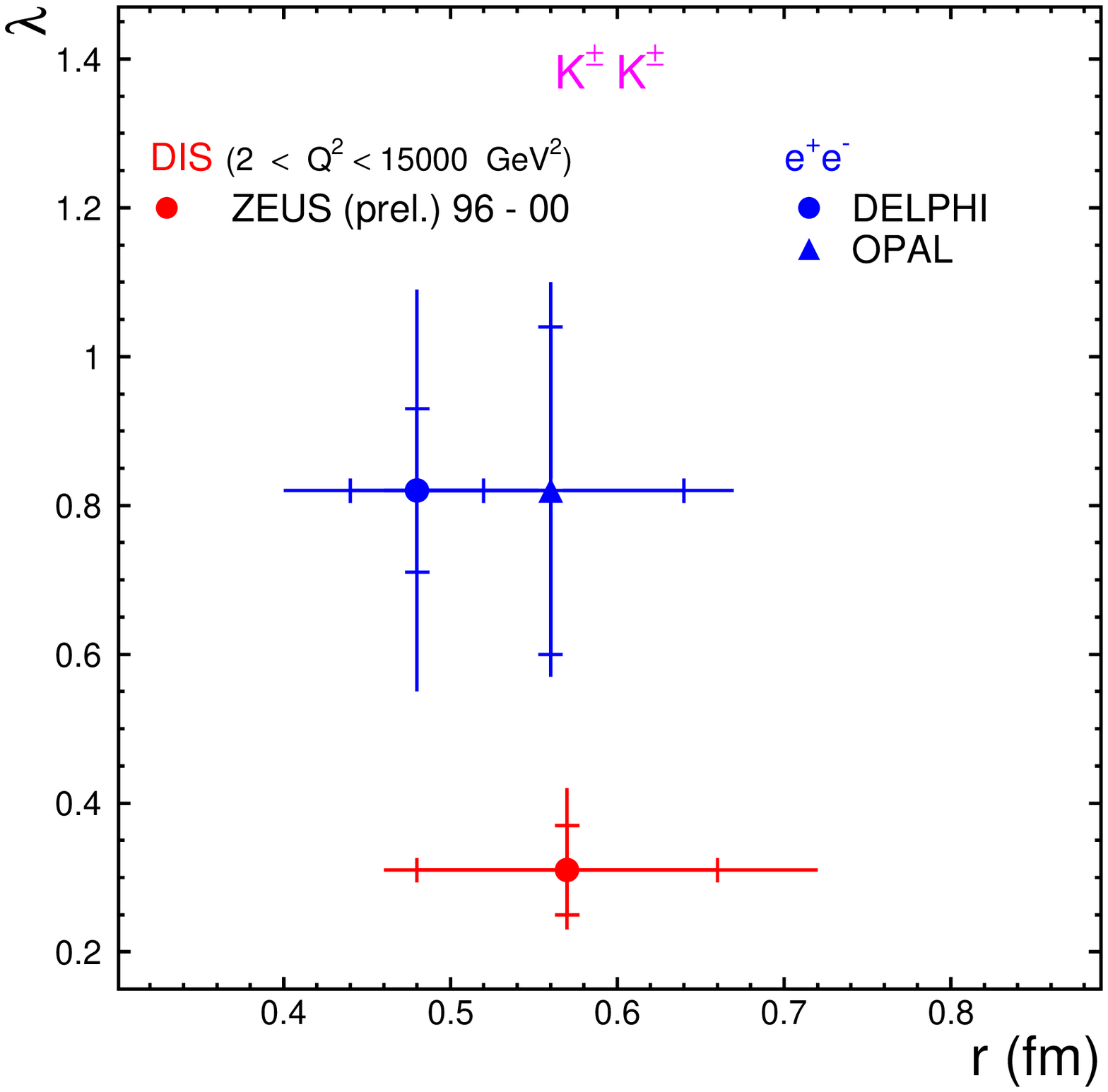}
\includegraphics[width=2.4in, height=2.3in]{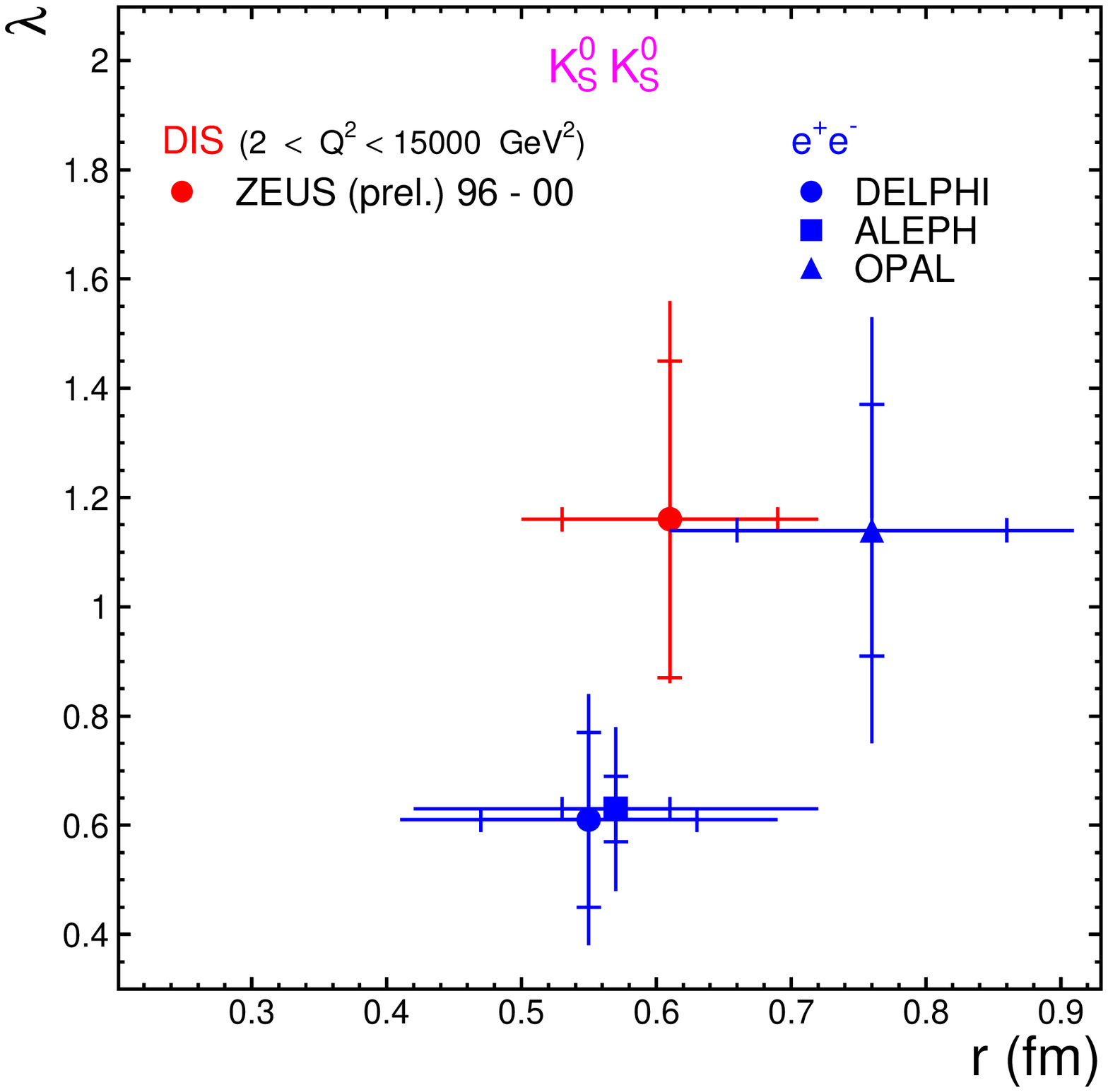}
\end{center}
\caption{Comparisons of the ZEUS and LEP BEC for charged and neutral kaons}
\label{fig2}
\end{figure} \\
Fig. 2 shows the results of BEC and the comparison of the ZEUS and LEP measurements. The measured radii are in good agreement with the LEP results while the $\lambda$ values are different. Different fragmentation processes in $e^+e^-$ and $ep$ may have effects on the $\lambda$ values. In addition, the decay of resonances may significantly increase the number of non-prompt kaons which will diminish the charged kaon correlation effect. In the case of neutral kaons, the value of $\lambda$ from ZEUS is larger than that measured by ALEPH\cite{ale} and DELPHI\cite{del} but close to the results from OPAL\cite{opa}. The difference can be explained by the different treatment of the  $f^0(980)$ decays, which were (not) removed in the case of ZEUS and OPAL (ALEPH and DELPHI).  
\section{Anti-Deuteron photoproduction}
Production of heavy stable particles such as deuterons, tritons and their anti-partners is not well understood in $ep$ collisions. In heavy ion collisions, their production is thought to be related to the so-called "thermal freeze-out" region\cite{fire,lett}. In the coalescence model\cite{coa1,coa2}, which describes heavy particle production, the production of a deuteron is related to the production of one free proton and one free neutron in the same reaction. The coalescence parameter, $B_2$, reflects the property of the initial "fireball" and is found to be inversely proportional to the size of the interaction region in heavy ion collisions\cite{fire}. \\
%\begin{figure}[h]
%\begin{center}
%\includegraphics[width=3.6in, height=2.6in, angle=0]{./d01.eps}
%\end{center}
%\caption{The observed dE/dx versus track momentum (in GeV) for the positively charged tracks (upper plot). The mass spectrum is shown in lower plot. The smooth curve is the Gaussian fit of the peak.}
%\label{fig3}
%\end{figure} \\
A measurement of anti-deuteron photoproduction was performed by the H1 Collaboration\cite{php}. In this study, heavy stable particles were identified using their most probable specific energy loss, dE/dx. After removing the beam-gas interaction and material backgrounds, a total of 45 $\pm$ 1.0 anti-deuterons were found in the H1 1996 data corresponding to an integrated luminosity of 5.53 $pb^{-1}$ at the average $\gamma p$ center-of-mass energy of 200 GeV. The kinematic range is 0.2 $<p_T/M<$ 0.7 and rapidity region $|y| <$0.4, where $p_T$ is the transverse momentum of the charged track and $M$ is the mass assigned to the track. In this kinematic region the total cross section is found to be 2.7 $\pm$ 0.5 nb. 
\begin{figure}[h]
\begin{center}
\includegraphics[width=2.4in, height=2.3in, angle=0]{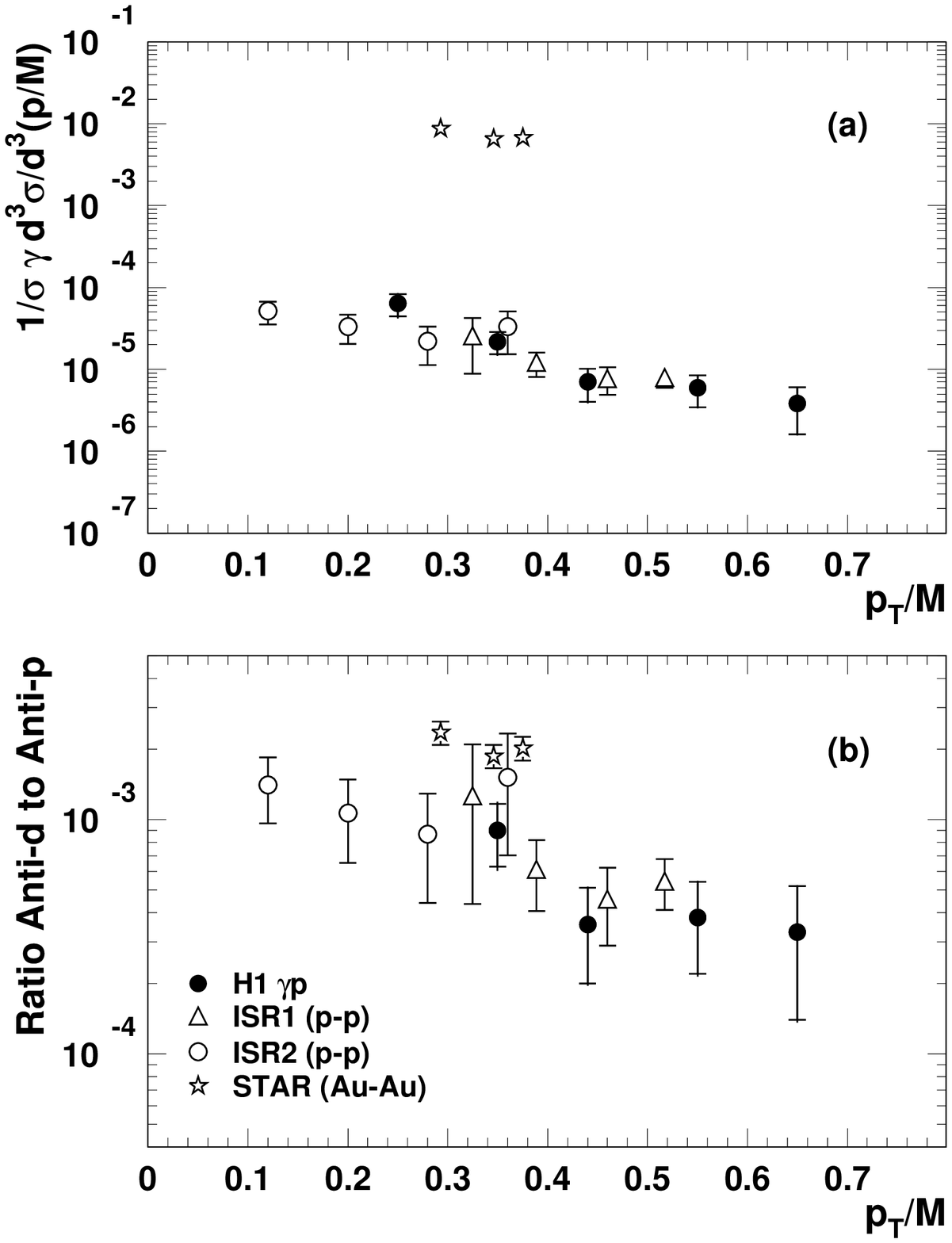}
\includegraphics[width=2.4in, height=2.2in,angle=0]{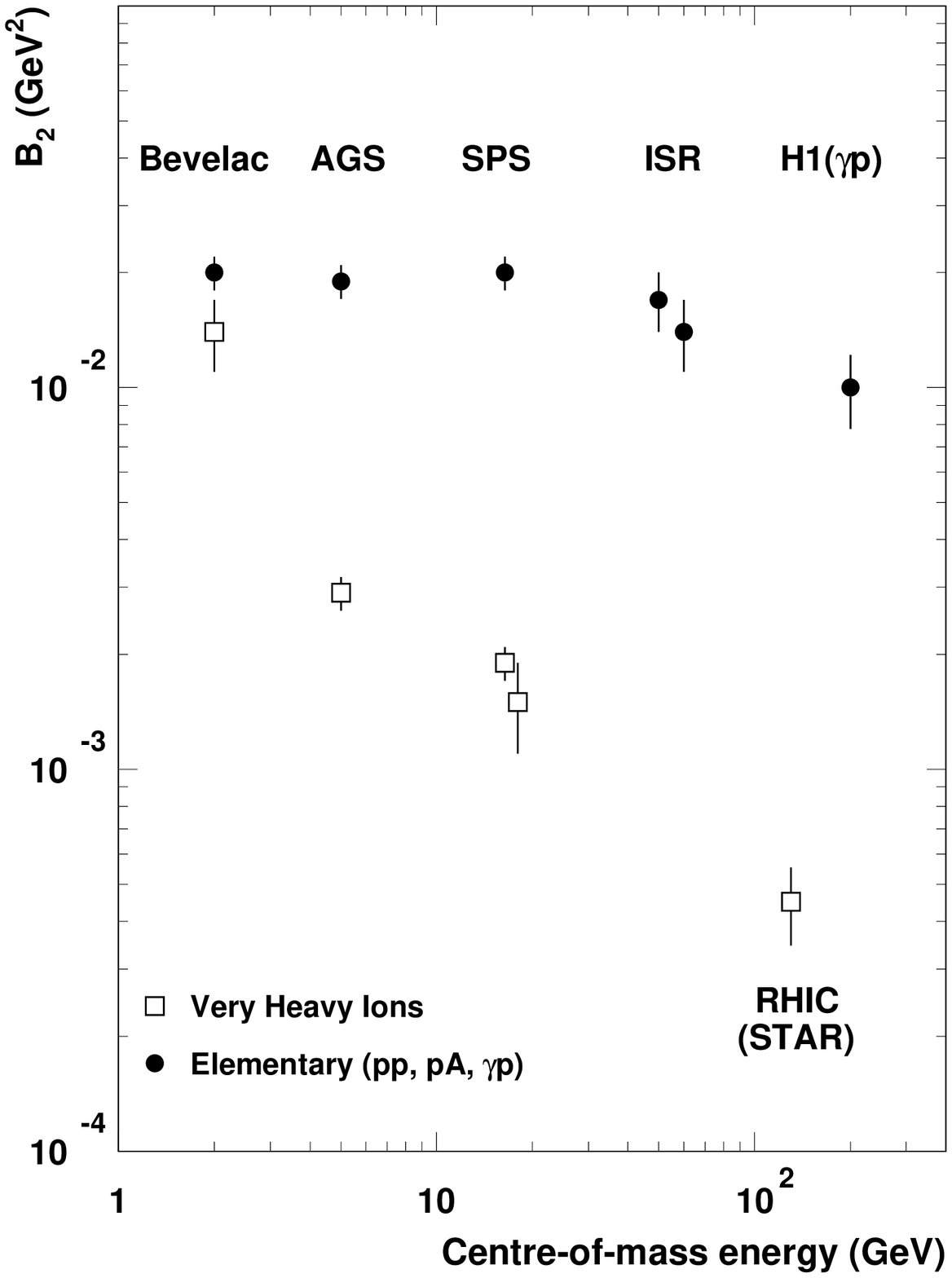}
\end{center}
\caption{The measured invariant cross section for inclusive $\bar{d}$ production (a) and $\bar{d}$ to $\bar{p}$ relative production ratio (b) as a function of $p_T/M$ compared with other experimental measurements. The $p_T/M$ dependence of the parameter $B_2$ (right plot).}
\label{fig3}
\end{figure} \\
Fig. 3 (a) shows that the normalised invariant $\bar{d}$ cross sections obtained in $\gamma p$ and $pp$ collisions are in good agreement but much lower than in the Au-Au experiment. However plot (b) shows that the $\bar{d}$ to $\bar{p}$ relative productions are similar in elementary particle collisions and in heavy ion collisions. The measured $B_2$ value in $\gamma p$ collisions is similar to the values in $pp$ and $pA$ at a lower center-of-mass energy, though more than an order of magnitude larger than in Au-Au collisions at comparable center-of-mass energies. This may be an indication that the coalescence model can describe heavy stable particle production well if the interaction volume of the thermal freeze-out in elementary particle collisions is much smaller than that in heavy ion collisions at a center-of-mass energy of 200 GeV.  
\section{Azimuthal asymmetry}
Measurement of the azimuthal distribution of the final state hadrons in semi-inclusive DIS processes provides a way to test pQCD predictions. It is of interest to investigate the azimuthal distribution in the hadronic center-of-mass (HCM) frame, where the azimuthal angle, $\phi$, is defined as the angle between the hadron production plane and the lepton scattering plane. The semi-inclusive DIS cross section can be expressed as:
\begin{center}
$\frac{d \sigma^{ep \rightarrow ehX}}{d \phi} = A + B cos(\phi) + C cos(2\phi) + D sin(\phi) + E sin(2\phi)$,
\end{center}
where $B$, $C$, $D$ and $E$ denote the asymmetry parameters. These parameters can be determined experimentally via calculating the 1st order moments of the respective trigonometrical functions of $\phi$:
\begin{center}
$\langle cos(\phi) \rangle = \frac{B}{2A}$ \hspace*{1cm}  $\langle cos(2\phi) \rangle = \frac{C}{2A}$ \\
$\langle sin(\phi) \rangle = \frac{D}{2A}$ \hspace*{1cm}  $\langle sin(2\phi) \rangle = \frac{E}{2A}$
\end{center}
The expressions above show that an azimuthal asymmetry exists only if the final hadrons have transverse momentum/energy (non-zero $\phi$). Hence the main processes which can give rise to azimuthal asymmetries are two-body processes like QCD Compton (QCDC) and boson-gluon fusion (BGF). The polarisation of the electron beam at HERA is also expected to result in non-zero asymmetries, since the spin of the electrons can be  transfered to the final hadronic states. The results reported here are based on the ZEUS 1995-1997 data with an integrated luminosity of 45 $pb^{-1}$ from unpolarised positron proton collisions.
\begin{figure}[h]
\begin{center}
\includegraphics[width=2.7in, height=2.6in]{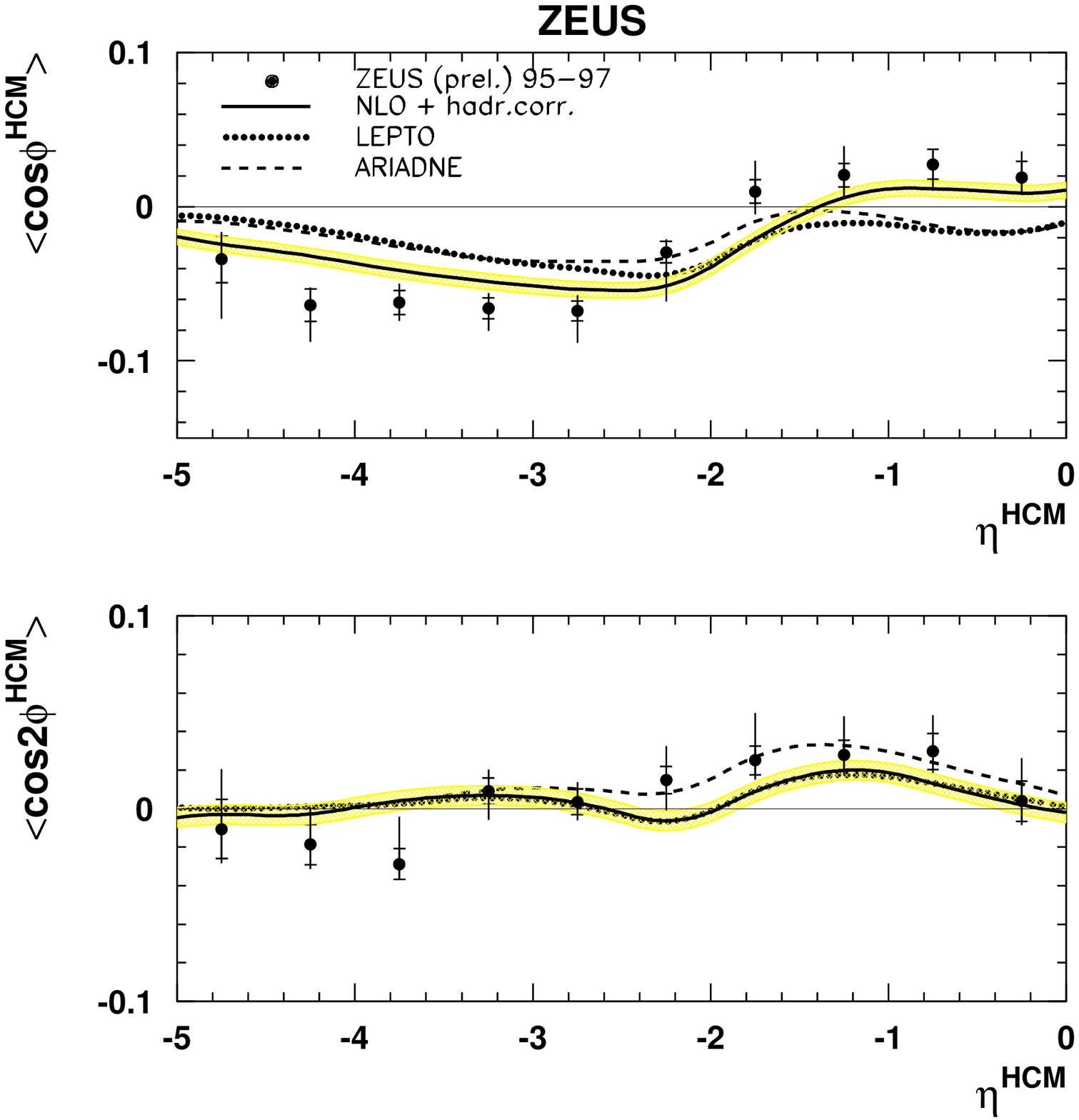}
\includegraphics[width=2.7in, height=2.6in]{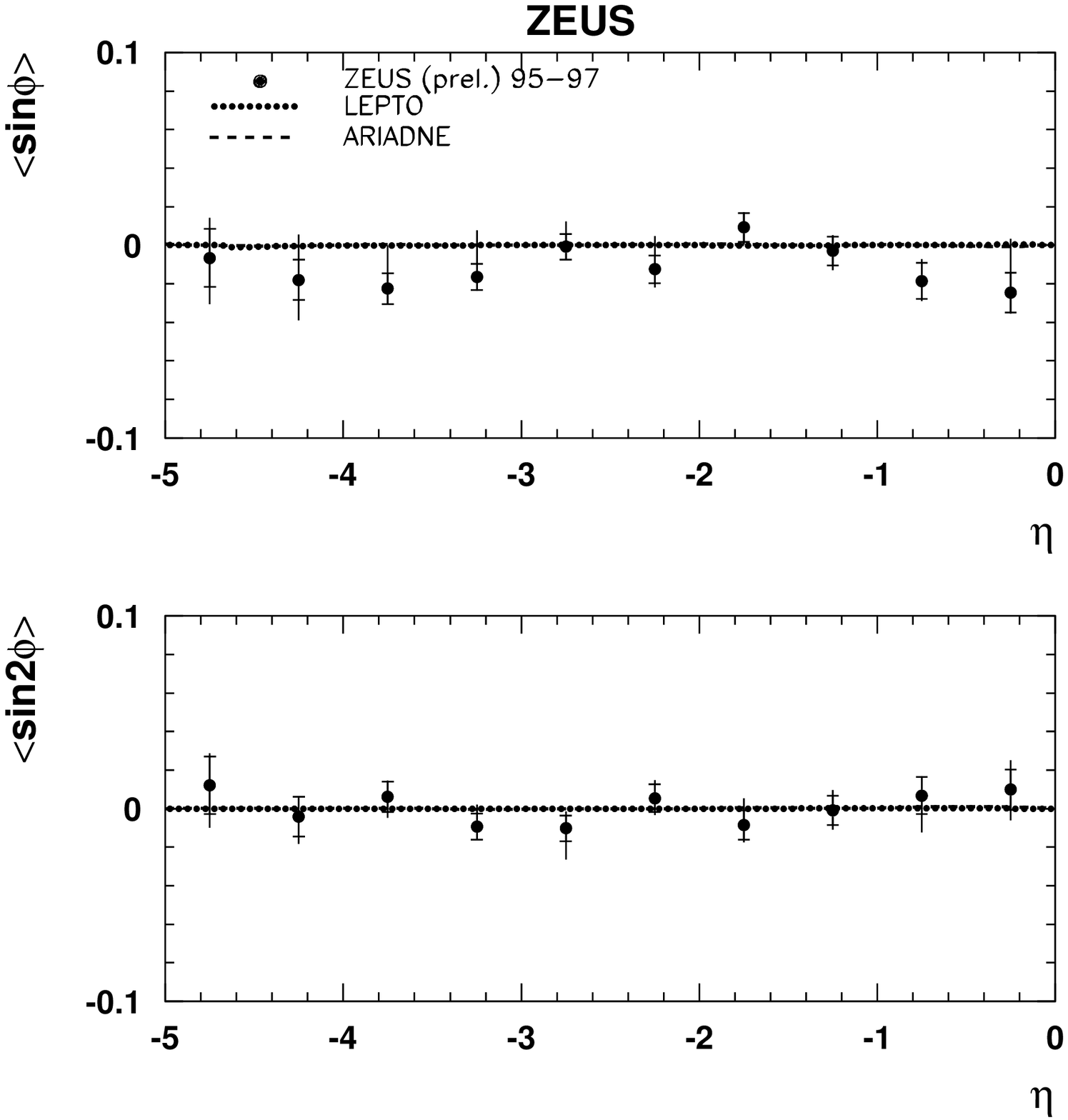}
\end{center}
\caption{The azimuthal asymmetry distribution $<cos\phi>$,  $<cos2\phi>$, $<sin\phi>$ and $<sin2\phi>$ are shown as a function of hadron pseudorapidity $\eta^{HCM}$ in the hadronic center-of-mass system.}
\label{fig4}
\end{figure}\\
The reconstruction of the hadronic final state is done using energy flow method \cite{flow,energy}. The main advantage of this method is that both calorimeter and tracking detector information are used so that the charged and neutral hadrons can be included in the investigation. In order to enhance contributions from hard partons, the direction of each final particle is weighted with its transverse energy. The kinematic region for this study is $ 100 < Q^2 <8000$ $\mathrm{GeV}^2$, $0.01 < x < 0.1$, $p^\mathrm{Lab}_\mathrm{T} > 150$ MeV.\\ 
The azimuthal asymmetry results for $<cos\phi>$,  $<cos2\phi>$, $<sin\phi>$ and $<sin2\phi>$ are shown in Fig. 4. The measured $<cos\phi>$ value is negative for $\eta < -2$ but becomes positive for larger $\eta$. It does not agree with the leading-order (LO) predictions from LEPTO and ARIADNE. LO calculations give negative values over the whole measured $\eta$ region. Next-to-leading order (NLO) predictions are in a better agreement with data. The shaded band is the theoretical uncertainties due to renormalisation and factorisation scales. For $<cos2\phi>$, the NLO and LO both match the data well. The mean value of $<sin\phi>$ and $<sin2\phi>$ are shown on the right hand side of Fig. 4. The measured mean values are consistent with zero and in good agreement with LO MC predictions. 
\section{Summary}
The strange particle production and correlations were investigated in deep inelastic scattering at HERA. Whereas the high $p_T$ region is well described by ARIADNE, the region of low $p_T$ is not. No significant $\Lambda$ to $\bar{\Lambda}$ asymmetry and $\Lambda$( $\bar{\Lambda}$) polarisations were observed. Bose-Einstein correlations of charged and neutral kaons were measured and compared to the LEP measurements. The photoproduction of $\bar{d}$ photoproduction has been studied and the measured coalescence parameter $B_2$ is found to be similar to that extracted from elementary particle collisions at a lower center-of-mass energy. The azimuthal distributions of the final hadrons in DIS were found to be reasonably well described by NLO predictions. \\ %LO MCs fail to reproduce the positive asymmetries observed in the pseudorapidity region above -2. \\

\end{document}